\documentstyle[twocolumn,aps,epsfig]{revtex}

\begin{document}
\voffset 1.8cm

\title{\LARGE \bf Inter-band B(E2) transition strengths \\ in odd-mass
heavy deformed nuclei}

\author{ 
C. E. Vargas$^1$\thanks{Electronic address: cvargas@fis.cinvestav.mx},
J. G. Hirsch$^2$\thanks{Electronic address: hirsch@nuclecu.unam.mx},
J. P. Draayer$^3$\thanks{Electronic address:
draayer@lsu.edu},\\
{\small $^1$  Departamento de F\'{\i}sica, Centro de Investigaci\'on y de
Estudios Avanzados del IPN,}\\
{\small Apartado Postal 14-740 M\'exico 07000 DF, M\'exico}\\
{\small $^2$  Instituto de Ciencias Nucleares, Universidad Nacional
Aut\'onoma de M\'exico,}\\
{\small Apartado Postal 70-543 M\'exico 04510 DF, M\'exico }\\
{\small $^3$  Department of Physics and Astronomy, Louisiana State
University,}\\
{\small Baton Rouge, LA 70803-4001, USA }
}

\date{\today}

\maketitle

\begin{abstract} {\bf Abstract}
Inter-band B(E2) transition strengths between different normal parity
bands in $^{163}$Dy and $^{165}$Er are described using the pseudo-SU(3)
model. The Hamiltonian includes Nilsson single-particle energies,
quadrupole-quadrupole and pairing interactions with fixed, parametrized
strengths, and three extra rotor terms used to fine tune  the energy
spectra. In addition to inter-band transitions, the energy spectra and
the ground state intra-band B(E2) strengths are reported. The results
show the pseudo-SU(3) shell model to be a powerful microscopic theory
for a description of the normal parity sector in heavy deformed odd-A
nuclei.

\end{abstract}

\vskip2pc

The pseudo-SU(3) model \cite{Hech69,Ari69,Rat73} has been used to describe
normal parity bands in heavy deformed nuclei. The scheme  takes full
advantage of the existence of pseudo-spin symmetry, which refers to the
fact that single-particle orbitals with $j = l - 1/2$ and $j = (l-2) +
1/2$ in the $\eta$ shell lie close in energy and can therefore be
labeled as pseudo-spin doublets with quantum numbers
$\tilde j = j$, $\tilde\eta = \eta -1$ and
$\tilde l = l - 1$. The origin of this symmetry has been traced back to
the relativistic Dirac equation \cite{Blo95,Gin97,Men98}.
In the simplest version of the pseudo-SU(3) model, the intruder level with
opposite parity in each major shell is removed from active consideration
and pseudo-orbital and pseudo-spin angular momenta are assigned to the
remaining single-particle states.

A fully microscopic description of low-energy bands in even-even and
odd-A nuclei has been developed using the pseudo-SU(3) model. The first
applications used pseudo-SU(3) as a dynamical symmetry, with a single
irreducible representation (irrep) of SU(3) describing the yrast band up
to the backbending regime \cite{Dra82}. A comparison of the quantum rotor
and microscopic SU(3) states \cite{Cas88} provided a classification of the
SU(3) irreps in terms of their transformation properties under $\pi$
rotations in the intrinsic frame \cite{Les87} and led to the construction
of a $K^2$ operator that plays a crucial role, for example, in
determining the excitation energy of the gamma band \cite{Naq90}. On the
computational side, the development of a computer code to calculate
reduced matrix elements of physical operators between different SU(3)
irreps \cite{Bah94} represented a breakthrough in the development of the
pseudo-SU(3) model. With this code in place it is possible to include
symmetry breaking terms in the interaction, such as pairing which is
known to represent important two-body correlations in low-energy
configurations. These correlations, while most important for near closed
shell nuclei, are also essential for an accurate determination of the
low-lying structure of strongly deformed systems.

Once a basic understanding of this overall structure was achieved, a
powerful shell-model theory for a description of normal parity states in
odd-mass  heavy deformed nuclei emerged. For example, the low-energy
spectra of several A=159 isotopes \cite{Var00}, as well as their B(E2)
intra-band transitions strengths \cite{Hir00}, have been successfully
described within the pseudo-SU(3) framework using a realistic Hamiltonian.

In the present letter we examine the ability of the pseudo-SU(3) model to
describe inter-band B(E2) transition strengths in the odd-A $^{163}$Dy and
$^{165}$Er nuclei. In addition, we examine predictions of the theory for
low-energy spectra and ground-state intra-band B(E2) transition strengths
in these nuclei. These rare-earth nuclei have one unpaired neutron. This
means the normal parity bands have negative parity. The results represent
a full implementation of a very ambitious program implied in first
applications of the pseudo-SU(3) model to odd-mass nuclei performed
nearly thirty years ago \cite{Rat73}.

Many-particle states of $n_\alpha$ active nucleons in a given normal
parity shell $\eta_\alpha$, $\alpha = \nu$ (neutrons) or $\pi$ (protons),
can be classified by the following group chain:

\begin{eqnarray}
~ \{ 1^{n^{N}_\alpha} \} ~~~~~~~ \{ \tilde{f}_\alpha \} ~~~~~~~\{ f_\alpha
\} ~\gamma_\alpha ~ (\lambda_\alpha , \mu_\alpha ) ~~~ \tilde{S}_\alpha
~~ K_\alpha  \nonumber \\
U(\Omega^N_\alpha ) \supset U(\Omega^N_\alpha / 2 ) \times U(2) \supset
SU(3) \times SU(2) \supset \nonumber \\
\tilde{L}_\alpha  ~~~~~~~~~~~~~~~~~~~~~ J^N_\alpha ~~~~ \nonumber \\
SO(3) \times SU(2) \supset SU_J(2),
\label{eq:chains}
\end{eqnarray}

\noindent where above each group the quantum numbers that characterize its
irreps are given and $\gamma_\alpha$ and $K_\alpha$ are multiplicity
labels of the indicated reductions.

The most important configurations are those with highest spatial symmetry
\cite{Dra82,Var98}. This implies that $\tilde{S}_{\pi , \nu} = 0$ or
$1/2$, that is, only configurations with pseudo-spin zero for an even
number of nucleons or $1/2$ for an odd number are taken into
account. The basis is built by selecting those proton and neutron SU(3)
irreps with the largest value of the second order Casimir operator $C_2$,
and coupling them to a total SU(3) irrep  with good angular momentum.
Details can  be found in previous publications \cite{Dra82,Var00,Hir00}.

The Hamiltonian includes spherical Nilsson single-particle terms for
the protons and neutrons ($H_{sp,\pi[\nu]}$), the quadrupole-quadrupole
($\tilde Q \cdot \tilde Q$) and pairing ($H_{pair,\pi[\nu]}$)
interactions, as well as three `rotor-like' terms which are diagonal in
the SU(3) basis:

\begin{eqnarray}
  H & = & H_{sp,\pi} + H_{sp,\nu} - \frac{1}{2}~ \chi~ \tilde Q \cdot
          \tilde Q - ~ G_\pi ~H_{pair,\pi} ~\label{eq:ham} \\
    &   & - ~G_\nu ~H_{pair,\nu} + ~a~ K_J^2~ +~ b~ J^2~ +~ A_{sym}~
          \tilde C_2 . \nonumber
\end{eqnarray}

\noindent The term proportional to $K_J^2$ breaks the SU(3) degeneracy
of the different K bands \cite{Naq90}, the $J^2$ term represents a small
correction to fine tune the moment of inertia, and the last term, $\tilde
C_2$, is introduced to distinguish between SU(3) irreps with $\lambda$ and
$\mu$  both even from the others with one or both odd \cite{Les87}. This
term serves to distinguishes between SU(3) irreps that belong to
$A$-type and $B_{\alpha}$-type ($\alpha = 1,2,3$) internal
configurations, respectively.

The Nilsson single-particle energies as well as the pairing and
quadrupole-quadrupole interaction strengths were taken from systematics
\cite{Rin79,Duf96}; only $a$, $b$ and $A_{sym}$ were used for fitting.
Parameter values are listed in Table I and are consistent with those
used in the description of neighboring even-even and odd-A nuclei
\cite{Var00,Hir00,Beu98,Dra00}.

\begin{table}
\begin{tabular}{ccccccc}
          &${\chi}$&$G_\pi$&$G_\nu$&   $a$   &  $b$   & $A_{sym}$ \\ \hline
$^{163}$Dy &0.00719 & 0.128 & 0.104 & -0.0400 &-0.0040 & 0.0016  \\
$^{165}$Er &0.00705 & 0.127 & 0.103 & -0.0050 &-0.0012 & 0.0011
\end{tabular}
\caption{Parameters used in the Hamiltonian
(\ref{eq:ham}).}
\label{para}
\end{table}

\begin{figure}
\vspace*{-1.7cm}
\hspace{-0.7cm}
\epsfxsize=8.6cm
\centerline{\epsfbox{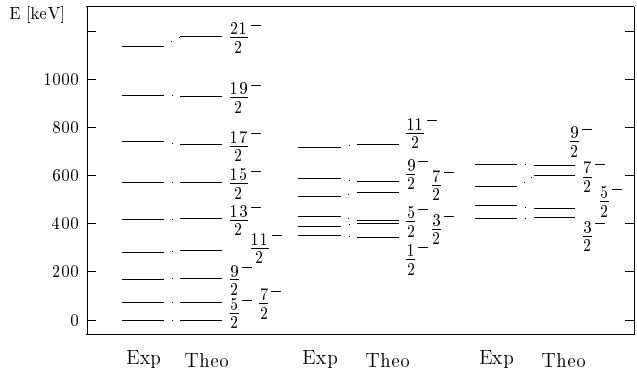}}\vspace{-5.2cm}
\caption{Energy spectra of $^{163}$Dy. `Exp' represents the experimental
results and `Theo' the calculated ones.}
\label{espdy163}\end{figure}

The E2 transition operator is given by \cite{Dra82}

\begin{equation}
Q_\mu = e_\pi Q_\pi + e_\nu Q_\nu \approx
e_\pi {\frac {\eta_\pi +1} {\eta_\pi}} \tilde Q_\pi +
e_\nu {\frac {\eta_\nu +1} {\eta_\nu}} \tilde Q_\nu ,
\end{equation}

\noindent with effective charges $e_\pi = 2.3$ and $e_\nu = 1.3$ for both
nuclei. These values are very similar to those used in a description of
the neighboring even-even \cite{Dra82,Beu98} and odd-A \cite{Hir00} nuclei
using the same model; they are larger than those used in standard
calculations of B(E2) strengths \cite{Rin79} due to the passive role
assigned to nucleons in the unique parity orbitals, whose contribution to
the quadrupole moments is parametrized in this way.

Figure \ref{espdy163} shows the calculated and experimental \cite{NNDC}
K = 5/2, 1/2, and 3/2 bands for $^{163}$Dy. The valence space included
10 protons and 9 neutrons in the normal parity pseudo oscillator shells
$\tilde{\eta}_\pi = 3$ and $\tilde{\eta}_\nu = 4$, respectively.
The agreement between theory and experiment is in general excellent.

\begin{figure}
\vspace*{-1.7cm}
\hspace{-0.7cm}
\epsfxsize=8.6cm
\centerline{\epsfbox{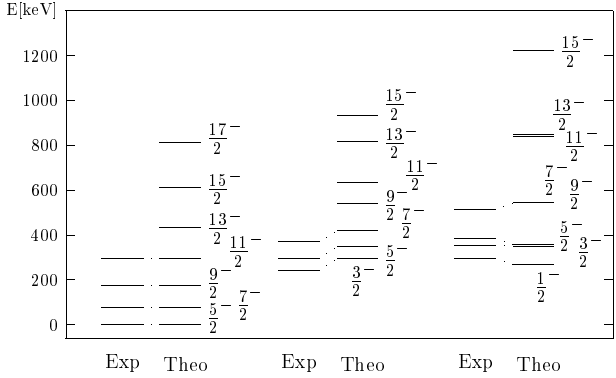}}\vspace{-5.2cm}
\caption{Energy spectra of $^{165}$Er. `Exp' represents the experimental
results and `Theo' the calculated ones.}
\label{esper165}\end{figure}

In Figure \ref{esper165} we present the low-lying energy spectra of
$^{165}$Er, including the K = 5/2, 3/2, and 1/2 bands built with twelve
protons in the normal parity $\tilde{\eta}_\pi = 3$ shell and nine
neutrons in the $\tilde{\eta}_\nu = 4$ shell.  In this case there
is also good agreement between experiment \cite{NNDC} and theory. The
model predicts states that have not been seen in the three lowest bands
and shows staggering in the excited bands from a simple $J(J+1)$ rule
that is expected for a pure rotor model. Due to the lack of experimental
information regarding these excited states, it is not possible to confirm
the existence of this staggering.

\begin{figure}\vspace{-1.7cm}
\epsfxsize=8cm
\centerline{\epsfbox{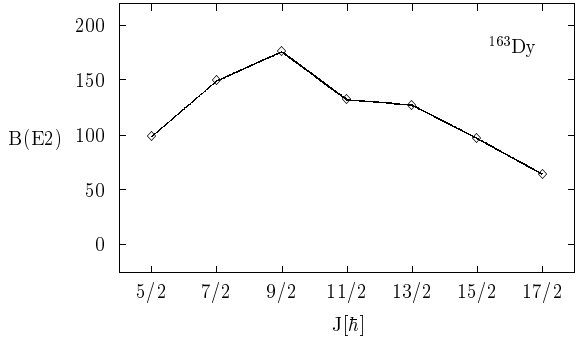}}\vspace{-5cm}
\caption{Theoretical B(E2;$J^-\rightarrow(J+2)^-$) trends in
$^{163}$Dy [$e^2b^2\times10^{-2}$].}
\label{be2dy163}\end{figure}

In Figure \ref{be2dy163} the intra-band B(E2) transition strengths of
the ground state band in $^{163}$Dy are shown.
Experimental values for intra-band B(E2) strengths in $^{163}$Dy are
reported for only three transitions: 5/2 ($\rightarrow$ 9/2), 97 $\pm$
18; 9/2 ($\rightarrow$ 13/2), 148 $\pm$ 59; and 11/2 ($\rightarrow$ 15/2),
205 $\pm$ 49 $e^2b^2\times10^{-2}$. Theoretical predictions for the 5/2
and 9/2 transitions are in good agreement with experimental values, but
the 11/2 prediction falls outside the error bars. For the
B(E2;$J^-\rightarrow(J+1)^-$) transition strengths, there is also good
agreement for all but the first 5/2 $\rightarrow$ 7/2 transition, which
theory overestimates.

\begin{figure}[t]
\vspace*{-1.7cm}
\epsfxsize=8cm
\centerline{\epsfbox{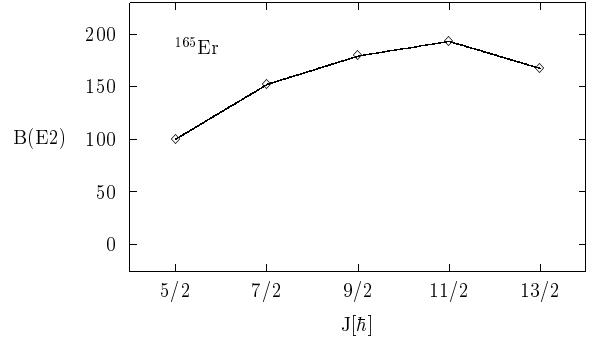}}\vspace{-5.2cm}
\caption{Theoretical B(E2;$J^-\rightarrow(J+2)^-$) B(E2) trends in
$^{165}$Er [$e^2b^2\times10^{-2}$].}
\label{be2er165}\end{figure}

The B(E2) values between states belonging the ground state band of
$^{165}$Er are shown in Figure \ref{be2er165}. There are no experimental
values reported for these transitions. Inter-band B(E2) transition
strengths for both nuclei are shown in Table \ref{tbe2inter}.

In what may be considered a defining characteristic of these nuclei,
inter-band transitions are typically two orders of magnitude smaller than
those between states in the same band. This characteristic feature of
deformed nuclei is reproduced by the model. A detailed analysis of
the wave functions yields a simple explanation. The wave functions of the
states belonging to the K = 5/2, 1/2 and 3/2 bands in $^{163}$Dy share
their main components. Their leading SU(3) component is between 50 and
68\% (30,6) and between 24 to 38\% (32,2), but the latter enter with a
different phases. These phases are responsible for the coherence of B(E2)
strengths within each band and at the same time for a cancellation among
the terms which leads to small numbers for inter-band transitions.

\begin{table}
\begin{tabular}{ccc}
           \multicolumn{3}{c}{B(E2) for $^{163}$Dy}       \\ \hline
    $J_i^- \rightarrow J_f^-$   & Theo. & Exp. \cite{NNDC} \\ \hline
$1/2^-_1 \rightarrow 5/2^-_1$  & 3.95     & $  4.0  \pm 0.8   $ \\
$3/2^-_1 \rightarrow 5/2^-_1$  & 2.16     & $  1.8  \pm 0.6   $ \\
$3/2^-_1 \rightarrow 7/2^-_1$  & 0.34   & $  3.7  \pm 1.6   $ \\
$5/2^-_2 \rightarrow 7/2^-_1$  & 2.11     & $  3.7  \pm 2.1   $ \\
$5/2^-_2 \rightarrow 9/2^-_1$  & 2.16     & $ 3.0   \pm 1.6   $ \\

                                &       &                 \\
                                &       &                 \\
           \multicolumn{3}{c}{B(E2) for $^{165}$Er}       \\ \hline
    $J_i^- \rightarrow J_f^-$   & Theo. & Exp. \cite{NNDC} \\ \hline
$1/2_2^- \rightarrow 3/2_1^-$  & 0.26  & $    >0.54    $ \\
$1/2_1^- \rightarrow 5/2_1^-$  & 1.83  & $1.08 \pm 0.13$ \\
$1/2_2^- \rightarrow 5/2_1^-$  & 0.54  & $    >0.05    $ \\
$3/2_1^- \rightarrow 7/2_1^-$  & 1.22  & $0.53 \pm 0.12$ \\
$3/2_2^- \rightarrow 7/2_1^-$  & 1.83  & $2.09 \pm 0.38$ \\

                                &       &
\end{tabular}
\caption{Theoretical and experimental inter-band B(E2)
transition strengths for $^{163}$Dy and $^{165}$Er in [$e^2 b^2 \times
10^{-2}$].}
\label{tbe2inter}
\end{table}

As shown in Table \ref{tbe2inter}, the agreement between experimental
\cite{NNDC} and theoretical B(E2) values for five transitions in
$^{163}$Dy is quite remarkable. For the inter-band B(E2) transition
strengths in $^{165}$Er, the situation is somewhat different.
Specifically, in addition to the relative phase changes noted for the
$^{163}$Dy case, in $^{165}$Er the main SU(3) components of the wave
function varies from one band to the next. For example, for the
transition $1/2^-_1 \rightarrow 5/2^-_1$,  the $1/2^-_1$ state is 97\%
(32,2) and 2\% of (31,4), while the $5/2^-_1$ is comprised of 67 and 31\%
of these irreps, respectively. So even though both states have the (32,2)
irrep as its main component, it enters in different amounts. In addition,
the phases of other components are such as to make the B(E2)
value very small. For $3/2^-_1 \rightarrow 5/2^-_1$ and $3/2^-_1
\rightarrow 7/2^-_1$ transitions, the SU(3) components of the states
involved are nearly the same, ranging from 23 to 31\% for the (31,4)
irrep and from 67 to 76\% for the (32,2) irrep. In these transitions,
differences in phases is the main effect yielding the orthogonality of
the wave functions, which means small inter-band B(E2) value.

It has been shown that normal parity bands in heavy deformed nuclei
can be described quantitatively using the pseudo-SU(3) model. This
coupling scheme, which was applied in a previous study to the low-energy
spectra \cite{Var00} and intra-band \cite{Hir00} B(E2) transition
strengths in odd-mass heavy deformed nuclei, was used to study
inter-band transition strengths. The agreement found with known
experimental values is good.
This shows that the pseudo-SU(3) scheme is a very creditable theory
for achieving a microscopic description of odd-mass deformed rare-earth
nuclei. The ability of the model to describe complementary properties of
odd-mass heavy deformed nuclei such as $g$ factors, the scissors mode and
beta decay transitions will be explored in the future.

This work was supported in part by Conacyt (M\'exico) and the National
Science Foundation under Grant PHY-9970769 and Cooperative Agreement
EPS-9720652 that includes matching from the Louisiana Board of Regents
Support Fund.


\end{document}